\documentstyle[preprint,prc,aps,epsf]{revtex}
\begin{document}
\tightenlines
\preprint{GSI-Preprint-2000-49, November 2000}
\title{Shadowing Effects on Vector Boson Production}

\author{R. Vogt}
 
\address{{
GSI, Planckstrasse 1, D-64291 Darmstadt\break
Nuclear Science Division, Lawrence Berkeley National Laboratory, 
Berkeley, CA
94720, USA\break 
Physics Department, University of California, Davis, CA 95616, USA}\break}
 
\vskip .25 in
\maketitle
\begin{abstract}

We explore how nuclear modifications to the nucleon structure functions,
shadowing, affect massive gauge boson production in heavy ion collisions at
different impact parameters.  We calculate the dependence of 
$Z^0$, $W^+$ and $W^-$ production on rapidity and impact parameter to
next-to-leading order in
Pb+Pb collisions at 5.5 TeV/nucleon to study 
quark shadowing at high $Q^2$.  We also compare our Pb+Pb results
to the $pp$ rapidity distributions at 14 TeV.
\end{abstract}

The heavy ion collisions at the LHC will be a rich testing ground for hard 
processes which should dominate particle production \cite{kje,spenlhc}.  One of
the most promising signatures of quark-gluon plasma production at the CERN SPS
is $J/\psi$ suppression \cite{rvrev,hsrev} which has been compared to the 
Drell-Yan continuum in the lepton pair mass range $2.9 < m < 4.5$ GeV
\cite{na50}.  
Both $J/\psi$ and Drell-Yan pair production
are calculable in perturbative QCD.  At the LHC, quarkonium suppression will be
difficult to compare to the dilepton continuum due to contributions from $c
\overline c$ and $b \overline b$ decays which have large uncertainties in
nuclear collisions \cite{LinVogt}.  Since the low mass dilepton continuum is
expected to be dominated by $b \overline b$ decays, the $Z^0$ was suggested as 
an alternative reference process for quarkonium suppression at the LHC 
\cite{gunvogt,kvatz}.  There are two difficulties with using the $Z^0$ as a
baseline for quarkonium suppression:  
the large mass differences, $m_{Z^0} \gg m_\Upsilon, m_{J/\psi}$,
and the difference in production mechanisms, predominantly $q \overline q$
for the $Z^0$ and $gg$ for quarkonium.  Both these differences are important
as far as nuclear effects are concerned.  However, the 
differences that reduce the value of the $Z^0$ as a baseline process are the 
same that make it an interesting object of study itself---the $Z^0$ 
provides a unique
opportunity to study the modifications of the quark distributions in the
nucleus at high $Q^2$.  Therefore, in this paper
we examine the possible effects of this 
shadowing on $Z^0$ production as well as
$W^+$ and $W^-$ production which are also quark dominated.  The impact 
parameter dependence of the shadowing effect will also be discussed.

We further address the issue of how to measure the shadowing effect.  Since
isospin will play an important role in quark-dominated processes, the 
comparison between Pb+Pb interactions with and without shadowing is less useful
than in gluon-dominated processes such as heavy quark production 
\cite{spenprl}.  In addition, the first, best, $pp$ data will be at the maximum
LHC energy of 14 TeV.  Therefore we will also present the predicted rapidity
distributions in Pb+Pb collisions at 5.5 TeV/nucleon, the $pp$ distributions
at 14 TeV, and the Pb+Pb/$pp$ ratios as a function of rapidity 
at the two energies.

The electroweak production and decay channels of the massive vector bosons
make them excellent candidates for shadowing studies since no hadronic
final-state rescattering is possible.  The $Z^0$ itself, with a 3.37\%
branching ratio to lepton pairs, will be easily observable by reconstructing
the peak in the dilepton spectra.  
Full reconstruction of the leptonic $W^\pm$ decays, $W^\pm
\rightarrow l^\pm \nu$, is not possible 
due to the missing energy given to the undetected neutrino but charged leptons
with momenta greater than 40 GeV should be prominent.  This decay
channel has been
used at the Tevatron to measure the asymmetry between $W^+$ and $W^-$ 
production since the asymmetry is sensitive to
the down to up quark ratio in the proton at intermediate values of $x$ 
and high $Q^2$ \cite{CDFdat}.  If the charged leptons from $W^\pm$ decays can
be identified in heavy ion collisions, such asymmetry measurements may also be
employed at the LHC to reduce systematic uncertainties 
and obtain a more meaningful 
determination of the $Q^2$ dependence of quark shadowing in the nucleus.

The next-to-leading order, NLO, cross section per nucleon for nuclei
$A$ and $B$ colliding at impact parameter $b$ and producing a vector
boson $V$ with mass $m$ at scale $Q$ is
\begin{eqnarray} \frac{1}{AB} \frac{d\sigma^V_{AB}}{dy
d^2b d^2r} & = & H_{ij}^V \int \,dz \,dz' dx_1\, dx_2 \,dx \, \delta
\bigg(\frac{m_V^2}{s} - x x_1 x_2 \bigg) 
\delta \bigg( y - \frac{1}{2} \ln \bigg( \frac{x_1}{x_2} \bigg) \bigg) 
\label{sigmajpsi} \\
&   & \mbox{} \times
\bigg\{ \sum_{i,j \in Q,\overline Q} C^{\rm ii}(q_i, \overline q_j)
\Delta_{q \overline q}(x) F_{q_i}^A(x_1,Q^2,\vec{r},z) F_{\overline 
q_j}^B(x_2,Q^2,\vec{b} - \vec{r},z') \nonumber \\
&   & \mbox{} +
\sum_{i,k \in Q, \overline Q} C^{\rm if}(q_i, q_k) \Delta_{qg}(x) \bigg[
 F_{q_i}^A(x_1,Q^2,\vec{r},z) F_g^B(x_2,Q^2,\vec{b} - \vec{r},z') \nonumber \\
&   & \mbox{} + F_g^A(x_1,Q^2,\vec{r},z) 
F_{q_j}^B(x_2,Q^2,\vec{b} - \vec{r},z') \bigg] \bigg\} \, \,  , \nonumber
\end{eqnarray}
where $H_{ij}^V$ is proportional to the leading order, LO, 
partonic $ij \rightarrow V$ cross section and $Q = u,d,s$ and $c$.  
The matrices $C^{\rm ii}$ and
$C^{\rm if}$ contain information on the coupling of the various quark flavors
to boson $V$.  

We assume that the parton densities $F_i^A(x,Q^2,\vec{r},z)$ can be
factorized into $x$ and $Q^2$ independent nuclear
density distributions, position and nuclear-number independent nucleon parton
densities, and a shadowing function $S^i(A,x,Q^2,\vec{r},z)$ 
that describes the modification of
the nuclear structure functions in position and momentum space.  Thus
we have
\begin{eqnarray}
F_i^A(x,Q^2,\vec{r},z) & = & \rho_A(s) S^i(A,x,Q^2,\vec{r},z) 
f_i^N(x,Q^2) \\
F_j^B(x,Q^2,\vec{b} - \vec{r},z') & = & \rho_B(s') S^j(B,x,Q^2,\vec{b} - 
\vec{r},z') f_j^N(x,Q^2) \nonumber \,\, 
\end{eqnarray}
where $f^N_i(x,Q^2)$ is the density of parton $i$ in the nucleon
and the radial variables $s$ and $s'$ are $s = \sqrt{r^2 +
z^2}$ and $s'=\sqrt{|\vec{b}-\vec{r}|^2 +z'^{\, 2}}$.  In the absence
of nuclear modifications,
$S^i(A,x,Q^2,\vec{r},z)\equiv1$.  The nuclear density distribution is
described by a
Woods-Saxon parameterization,
\begin{equation}
\rho_A(s)= \rho_0 {1 + \omega(s/R_A)^2 \over 1 + \exp[(s-R_A)/d]} \,
\, .
\label{density}
\end{equation}
Electron scattering data \cite{Vvv} are used to fix 
the parameters $R_A$, $d$, $\omega$ and $\rho_0$.

Experiments \cite{Arn} have shown that the proton and neutron
structure functions are modified in the nucleus.  For
momentum fractions $x< 0.1$ and $0.3<x<0.7$, a depletion is observed
in a heavy nucleus relative to a light nucleus such as the deuteron.  
The low $x$, shadowing,
region and the larger $x$, EMC, region is bridged by an enhancement
at $0.1 < x < 0.3$ called antishadowing.  Here we refer to the modification
over the entire $x$ range as `shadowing' unless otherwise noted.
Many theoretical explanations 
have been proposed, typically for only part of the $x$ range such as
the very
low $x$ or EMC regions.  However, as none of the models can describe the
effect over all $x$ and $Q^2$, we rely on parameterizations of the nuclear 
modifications based on fits to data, as described later.

Most typical structure function measurements are insensitive to any spatial 
dependence and thus average over the entire nuclear volume.  
One experiment using a
bubble chamber found that the structure function
does vary spatially but could not determine the dependence on impact
parameter\cite{e745}.  In a nuclear collision, 
the impact parameter can be determined from the transverse
energy production.  The influence of the spatial dependence of shadowing
on transverse energy production has already been
considered\cite{spenlhc,Eskola2}. The effects of spatially
inhomogeneous shadowing on heavy quark \cite{spenprl,spenprc},
quarkonium, and Drell-Yan \cite{spenlhc,spenpsi} production in heavy
ion collisions has also been discussed previously.

We now describe the NLO cross section in Eq.~(\ref{sigmajpsi}) in more detail.
The functions $H_{ij}^V$ are rather simple \cite{zwrefs}:
\begin{eqnarray}
H_{ij}^{Z^0} & = & \frac{8\pi}{3} \frac{G_F}{\sqrt{2}} [(g_V^i)^2 
+ (g_A^i)^2] \frac{m_Z^2}{s} \label{zpart} \\
H_{ij}^{W^\pm} & = & \frac{2\pi}{3} \frac{G_F}{\sqrt{2}} \frac{m_W^2}{s} 
\label{wpmpart} 
\end{eqnarray}
where $G_F = 1.16639 \times 10^{-5}$ GeV$^2$, $m_Z = 91.187$ GeV, and $m_W = 
80.41$ GeV.  For $Z^0$ production by a given flavor $i$ with charge $e_i$, 
the sum of the 
squared vector and axial vector couplings is $(g_V^i)^2 + (g_A^i)^2 = 
(1/8)(1 - 4 | e_i | x_W
+ 8 e_i^2 x_W^2)$ where $x_W = \sin^2 \theta_W = 1 - m_W^2/m_Z^2$. 

The functions $\Delta_{ij}(x)$ are 
universal for all vector bosons, including virtual photons produced in the
Drell-Yan process \cite{zwrefs}.  We work in the
${\overline {\rm MS}}$ scheme.  The NLO correction to the $q \overline q$
channel includes the contributions from soft and virtual gluons as well as hard
gluons from the process $q \overline q \rightarrow V g$.  We have, up to NLO
\cite{zwrefs},
\begin{eqnarray} \Delta_{q \overline q}(x) & = & \delta(1-x) +
\frac{\alpha_s(Q^2)}{3\pi} \bigg\{ -4(1 + x) \ln\bigg(\frac{Q^2}{m_V^2} \bigg)
-8(1+x) \ln(1-x) - 4 \frac{1+x^2}{1-x} \ln x  \label{delqqbar} \\
&   & \mbox{} + \delta(1-x) \bigg[ 6 \ln \bigg(\frac{Q^2}{m_V^2} \bigg) + 8
\zeta(2) - 16 \bigg] + 8 \bigg[ \frac{1}{1-x} \bigg]_+  
\ln\bigg(\frac{Q^2}{m_V^2} \bigg)
+ 16 \bigg[ \frac{\ln(1-x)}{1-x} \bigg]_+ \bigg\} \, \, . \nonumber
\end{eqnarray}
The first delta function is the LO contribution while the NLO contribution is
proportional to $\alpha_s(Q^2)$.  At NLO $\alpha_s(Q^2)$ 
is calculated to two loops
with $n_f = 5$ active flavors.  The last three terms are the soft and virtual
gluon contributions.  
The general integral of the 'plus' functions in the last two terms is 
\cite{hsv}
\begin{eqnarray}
\int_a^1 dx f(x) \bigg[ \frac{\ln^i (1-x)}{1-x} \bigg]_+ = \int_a^1 dx 
\frac{f(x)
- f(1)}{1-x} \ln^i(1-x) + \frac{f(1)}{i+1} \ln^{i+1}(1-a) \, \, .
\label{plus}
\end{eqnarray}
The quark-gluon contribution only appears at ${\cal O}(\alpha_s)$ through
the real correction $qg \rightarrow qV$.  At this order \cite{zwrefs}, 
\begin{eqnarray}
\Delta_{qg}(x) = \frac{\alpha_s(Q^2)}{8\pi} \bigg\{ 2(1 + 2x^2 - 2x)\ln \bigg(
\frac{(1-x)^2Q^2}{xm_V^2} \bigg) + 1 - 7x^2 + 6x \bigg\} \label{delqg} \, \, .
\end{eqnarray}
For gauge boson production, we take $Q^2 = m_V^2$ 
and all terms proportional to 
$\ln(Q^2/m_V^2)$ drop out.  Using the delta functions in 
Eq.~(\ref{sigmajpsi}) we
find  $x_{1,2} = (m_V/\sqrt{xs}) 
\exp( \pm y)$.  As at LO, when $\Delta_{q \overline q}(x)$ is 
proportional to $\delta(1-x)$ in Eq.~(\ref{delqqbar}), $x_{1,2}' = 
(m_V/\sqrt{s}) \exp( \pm y)$.  The rather lengthy convolutions of the 
shadowing functions and parton distribution functions including isospin via the
proton and neutron numbers are given in the appendix.

We now define the coupling matrices in Eq.~(\ref{sigmajpsi}).  The
superscripts represent the initial (i) and final (f) state quarks or antiquarks
while the arguments indicate
the orientation of the quark line to which the boson is coupled \cite{zwrefs}.
The coupling matrices are fairly simple for $Z^0$ production: $C^{\rm ii}(q_i,
\overline q_j) = \delta_{ij}$ and  $C^{\rm if}(q_i, q_k) = \delta_{ik}$.
With $W^+$ and $W^-$ production, the couplings are elements of the CKM matrix.
They are nonzero for $C^{\rm ii}(q_k,\overline q_l)$ if $e_k + e_l = \pm 1$
and for $C^{\rm if}(q_k,q_l)$ if $e_k = \pm 1 + e_l$.  In both cases, they take
the values $| V_{q_k q_l} |^2$.  Following Hamberg {\it et al.} \cite{zwrefs},
we take $V_{ud} = \cos \theta_C \approx V_{cs}$ and $V_{us} =
\sin \theta_C \approx -V_{cd}$ with $\sin \theta_C \approx 0.22$.  

\begin{figure}[h]
\setlength{\epsfxsize=0.95\textwidth}
\setlength{\epsfysize=0.5\textheight}
\centerline{\epsffile{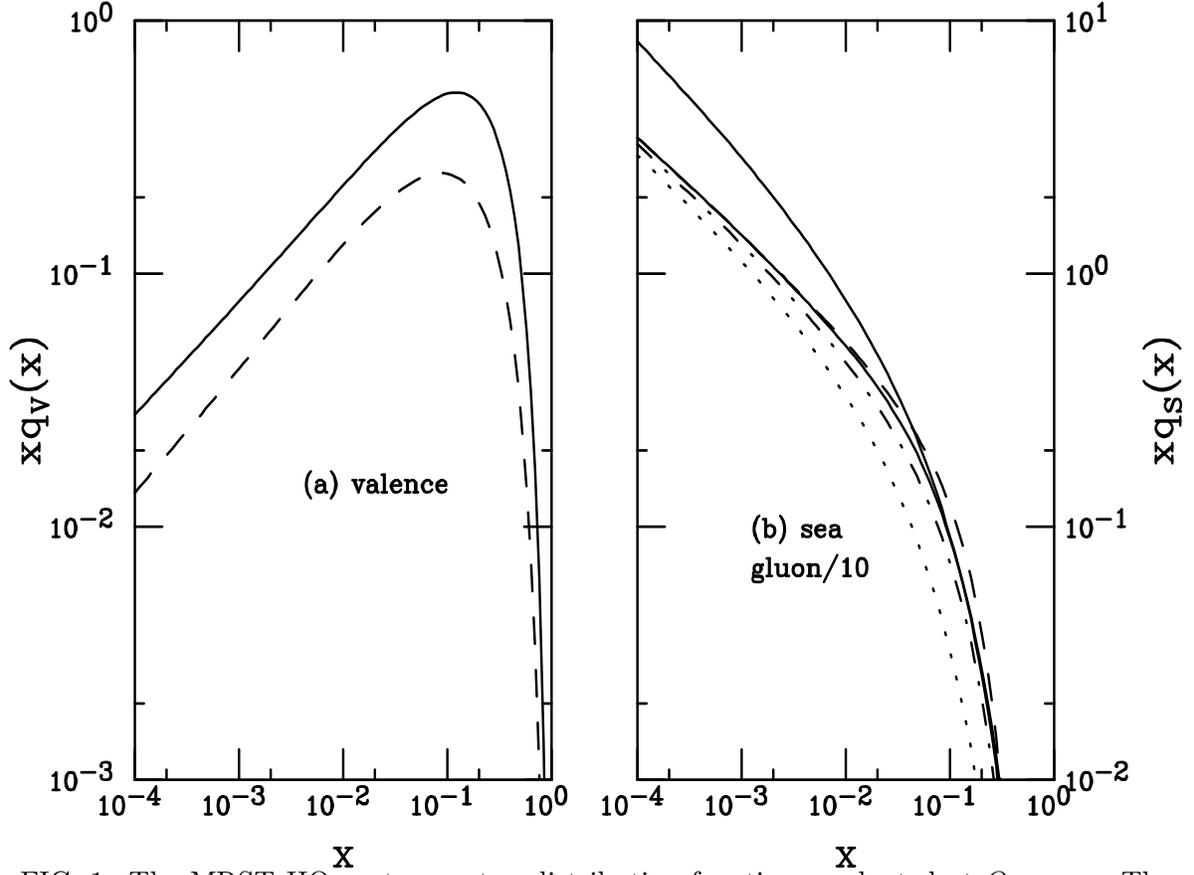}}
\caption{The MRST HO proton parton distribution functions evaluated at $Q =
m_{Z}$.  The up (solid) and down (dashed) valence distributions
are given in (a) while the up (lower solid), down (dashed), strange
(dot-dashed) and charm (dotted) sea quark distributions are shown in (b), along
with the gluon distribution (upper solid), 
reduced by a factor of 10 for comparison.
}
\label{pdfz}
\end{figure}

We use the MRST HO (central gluon)
\cite{mrsg} nucleon parton distributions in the ${\overline
{\rm MS}}$ scheme, shown evaluated
at $Q^2 = m_{Z}^2$ in Fig.~\ref{pdfz}.  The valence distributions are
somewhat larger than the corresponding sea quark distributions at $x\geq 0.1$ 
and extend to higher $x$ values.  The sea quarks dominate the valence 
quarks at $x \sim 10^{-4}$ by a factor of $\sim 100$.  Note also that
$f_{\overline d}^p$ is larger than $f_{\overline u}^p$ when $x > 0.01$.
The gluon distribution is shown at 1/10 of its magnitude. 
At low $x$, corresponding to large rapidity, the gluon density is
high.

The shadowing effect is studied with three parameterizations of the
average, homogeneous, shadowing, $S^i_k(A,x,Q^2)$ $\{k=1-3\}$, 
measured in nuclear
deep-inelastic scattering.  All the shadowing
parameterizations are obtained and evolved at leading order.  Since the
parameterizations are fit to ratios of heavy to light nuclei, the dependence
of the parameterizations on both the initial
parton densities and the order of the calculation should be weak.
The first, $S_1(A,x)$, assumes that
the quark, gluon and antiquark modifications are equivalent 
and includes no $Q^2$ evolution \cite{EQC}. The second,
$S_2^i(A,x,Q^2)$, has separate modifications for the valence quarks, sea quarks
and gluons and includes $Q^2$ evolution from $4 < Q^2 < 100$ GeV$^2$ 
\cite{KJE}.
The third parameterization, $S_3^i(A,x,Q^2)$, 
is based on the GRV LO
\cite{GRV} parton densities.  The ratios are evolved 
over $2.25 < Q^2 < 10^4$ GeV$^2$ \cite{EKRS3,EKRparam} assuming that
$S^{u_V}_3 = S^{d_V}_3$ and $S^{\overline u}_3 = S^{\overline d}_3$
while the more massive sea quarks are
evolved separately.  Both the $S_2$ and $S_3$ ratios are evolved to
higher $Q^2$ using the
DGLAP equations \cite{KJE,EKRS3,EKRparam}.  It was shown in Ref.~\cite{KJE}
that including recombination terms in the evolution, as in Ref.~\cite{MQiu}, 
did not have a large effect
on the shadowing ratios, particularly at the $x$ values probed here.  
The initial gluon
ratio in $S_3$ shows significant antishadowing for $0.1<x<0.3$
while the sea quark ratios are shadowed.  In contrast, $S_2$
has less gluon antishadowing and essentially no sea quark effect in
the same $x$ region.  Unfortunately, the $Q^2$ evolution of $S_2$
stops below the vector boson
mass, rendering it less valuable.  We show results with all three 
parameterizations because no nuclear DIS data is available at high $Q^2$.
Since the $S_3$ parameterization includes the most recent nuclear DIS
data and is evolved to scales compatible with the vector boson masses, 
the $S_3$ results should perhaps be favored. 

\begin{figure}[h]
\setlength{\epsfxsize=0.95\textwidth}
\setlength{\epsfysize=0.5\textheight}
\centerline{\epsffile{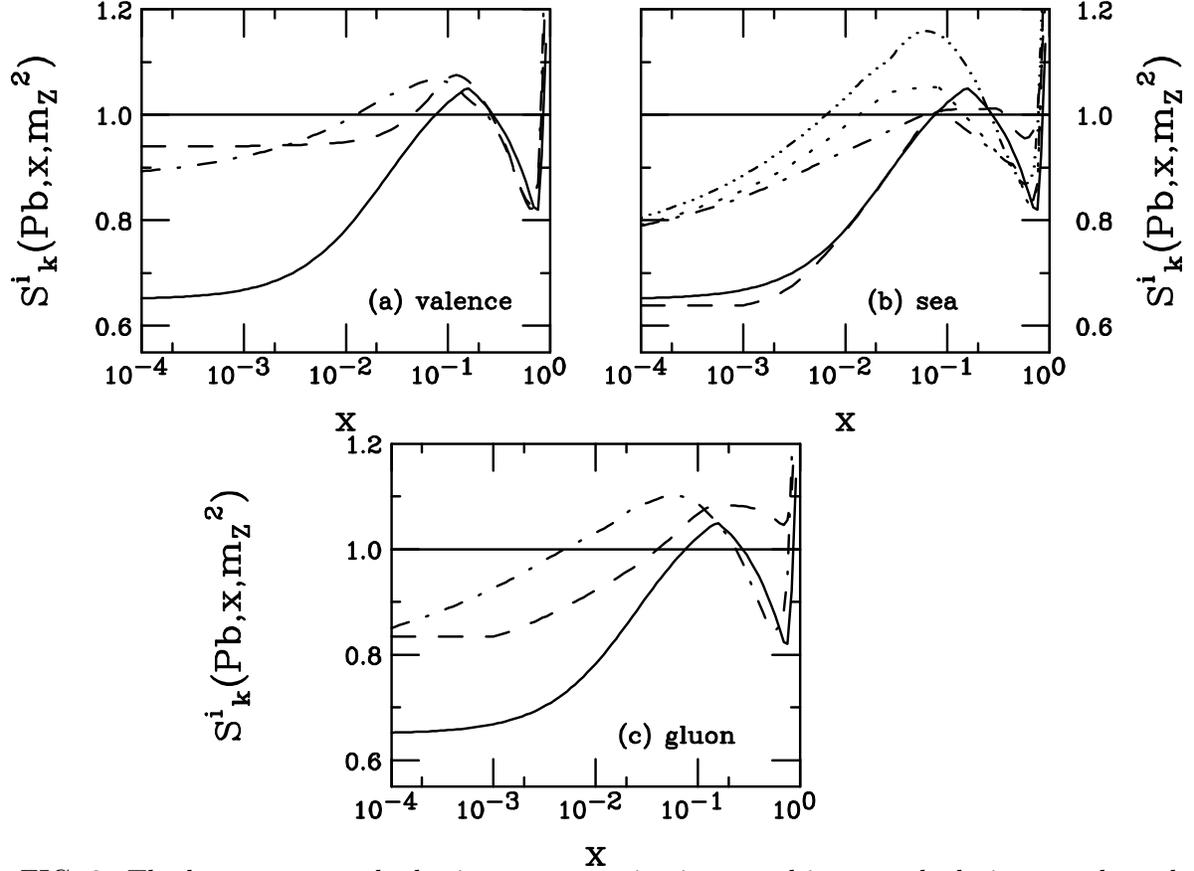}}
\caption{The homogeneous shadowing parameterizations used in our calculations,
evaluated at $Q^2 = m_{Z}^2$.  Valence shadowing is shown in (a) for the
$S_1$ (solid), $S_2^V$ (dashed), and $S_3$ (dot-dashed)
parameterizations.  Sea quark shadowing is shown in (b) for
$S_1$ (solid), $S_2^S$ (dashed), $S_3^{\overline u} = S_3^{\overline
d}$ (dot-dashed), $S_3^{\overline s}$ (dotted) and $S_3^{\overline c}$
(dot-dot-dot-dashed).  Gluon shadowing is shown in (c) for $S_1$ (solid),
$S_2^g$ (dashed) and $S_3^g$ (dot-dashed).
}
\label{shadrat}
\end{figure}

The shadowing ratios in a lead nucleus compared to a proton are shown in
Fig.~\ref{shadrat}.  The effects of shadowing on the valence quarks is 
strongest with $S_1$ since all quarks are treated equally.  The $S_2$
and $S_3$ valence ratios are rather similar in magnitude although the 
anti-shadowing range is broadest for $S_3$, $0.01 < x< 0.3$ and the $S_3$
ratio is lower than the $S_2$ ratio at low $x$.  The $S_1$ and $S_2$ sea quark 
ratios are very similar when $x < 0.1$.  Then the $S_2$ ratio is 
essentially unity until $x > 0.3$.  The $S_3$ ratios are all larger than the
$S_1$ and $S_2$ ratios, even at small $x$, due to evolution.
It is most interesting to note the 
difference between the light and strange sea ratios in the $S_3$ 
parameterization.  The ratios $S^{\overline u}_3$ and $S^{\overline d}_3$
show no antishadowing effect but instead decrease when $x > 0.1$ while 
$S^{\overline s}_3$ and $S^{\overline c}_3$ are typically larger over all $x$ 
and are
antishadowed when $0.01 < x<0.2$.  The antishadowing in the charm
distribution is larger even than for the gluons.
Since all $S_3$ sea ratios
are equivalent at $Q^2 = 2.25$ GeV$^2$, the difference is solely the effect of 
evolution.  Note also that the gluon shadowing ratios are typically larger than
the large $Q^2$ sea quark ratios over all $x$.  At $Q^2 = m_Z^2$ the strong
antishadowing in $S^g_3$ has essentially disappeared and is no larger than
that of $S_2^g$ although the antishadowing region of $S_3^g$ is broader, from
$0.005 < x < 0.2$.

Nuclear shadowing
should depend on spatial position of the partons in the nucleus
as well as on their momentum.  Most models predict some form of spatial
dependence, according to the origin of the shadowing effect.  Typically, the
spatial dependence can be expected to take two forms, either proportional to
the local nuclear density, Eq.~(\ref{density}), or the path length of the
parton through the nucleus.  Both will be discussed below.

When the parton density is high, partons in one nucleon can
interact with those in neighboring nucleons, recombining to lower the parton
density \cite{hot}.  In this case
shadowing is proportional to the local nuclear density
\cite{spenprl,spenprc}.  Then
\begin{eqnarray}
S^i_{k \, \rm WS} =
S^i_k(A,x,Q^2,\vec{r},z) & = & 1 + N_{\rm WS}
[S^i_k(A,x,Q^2) - 1] \frac{\rho_A(s)}{\rho_0} \label{wsparam} \, \, ,
\end{eqnarray}
where $N_{\rm WS}$ is a normalization constant chosen so that $(1/A) \int 
d^3s \rho_A(s)
S^i_{k \, \rm WS} = S^i_k$. At large radii, $s \gg R_A$, the medium
modifications weaken and the nucleons
behave as though they were free.  At the center of the nucleus, the
modifications are larger than the average value determined from nuclear DIS.

It has also been suggested that shadowing stems from multiple interactions
of the incident parton\cite{ayala}. In this picture, parton-parton 
interactions are longitudinally distributed over the coherence
length, $l_c=1/2m_Nx$, where $m_N$ is the nucleon mass\cite{ina}.  When
$x<0.016$, $l_c > R_A$ for all nuclei and the
interaction of the initial parton is delocalized over the entire nuclear
path, thus interacting coherently with all 
target partons along the distance $l_c$. For small $x$, shadowing
depends on the longitudinally-integrated nuclear density at transverse distance
$\vec{r}$ and the spatial dependence can then be parameterized as
\begin{eqnarray} S^i_{k \, \rho}(A,x,Q^2,\vec{r},z) = 1 + N_\rho 
(S^i_k(A,x,Q^2) - 1) \frac{\int dz
\rho_A(\vec{r},z)}{\int dz \rho_A(\vec{0},z)} 
\label{rhoparam} \, \,  \end{eqnarray}
where the normalization is again defined by $(1/A) \int d^2r dz \rho_A(s)
S^i_{k \, \rho} = S^i_k$ with $N_\rho > N_{\rm WS}$.
However, at large $x$,
$l_c\ll R_A$ and shadowing is again proportional to the local density so that
Eq.~(\ref{wsparam}) corresponds to the large $x$ limit of the multiple 
scattering formulation.

There are some problems  with implementing the multiple scattering
picture in nuclear collisions.
While traversing the formation length, both the initial- and final-state
partons may undergo multiple interactions, 
reducing the effective $l_c$, similar to the
Landau-Pomeranchuk-Migdal effect\cite{LPM}.  In addition, the idealized
picture of a single initial parton incident on a static nucleus is
inappropriate in heavy ion collisions since many interactions occur 
simultaneously, increasing the density in the path of the initial parton.  
A cascade approach cannot resolve the difficulty 
because non-local depictions of
these collisions are Lorentz frame dependent
\cite{bnl96}.  Finally, since the parton densities are distributed 
over an $x$-dependent distance, baryon number is not
locally conserved even if the valence quarks are considered to be fixed 
spatially.  Given the difficulties with the multiple interaction picture as 
well as those
of matching the two spatial dependencies according to $l_c$ at each
$x$, we only present specific results for the local
density model, Eq.~(\ref{wsparam}). 

Other mechanisms of shadowing effects in the EMC
region such as
nuclear binding \cite{close} and rescaling \cite{crr,cast}
have also been suggested but can
explain only part of the
observed effect \cite{li}. We note that these models would also
predict some spatial dependence.

\begin{figure}[htb]
\setlength{\epsfxsize=0.95\textwidth}
\setlength{\epsfysize=0.7\textheight}
\centerline{\epsffile{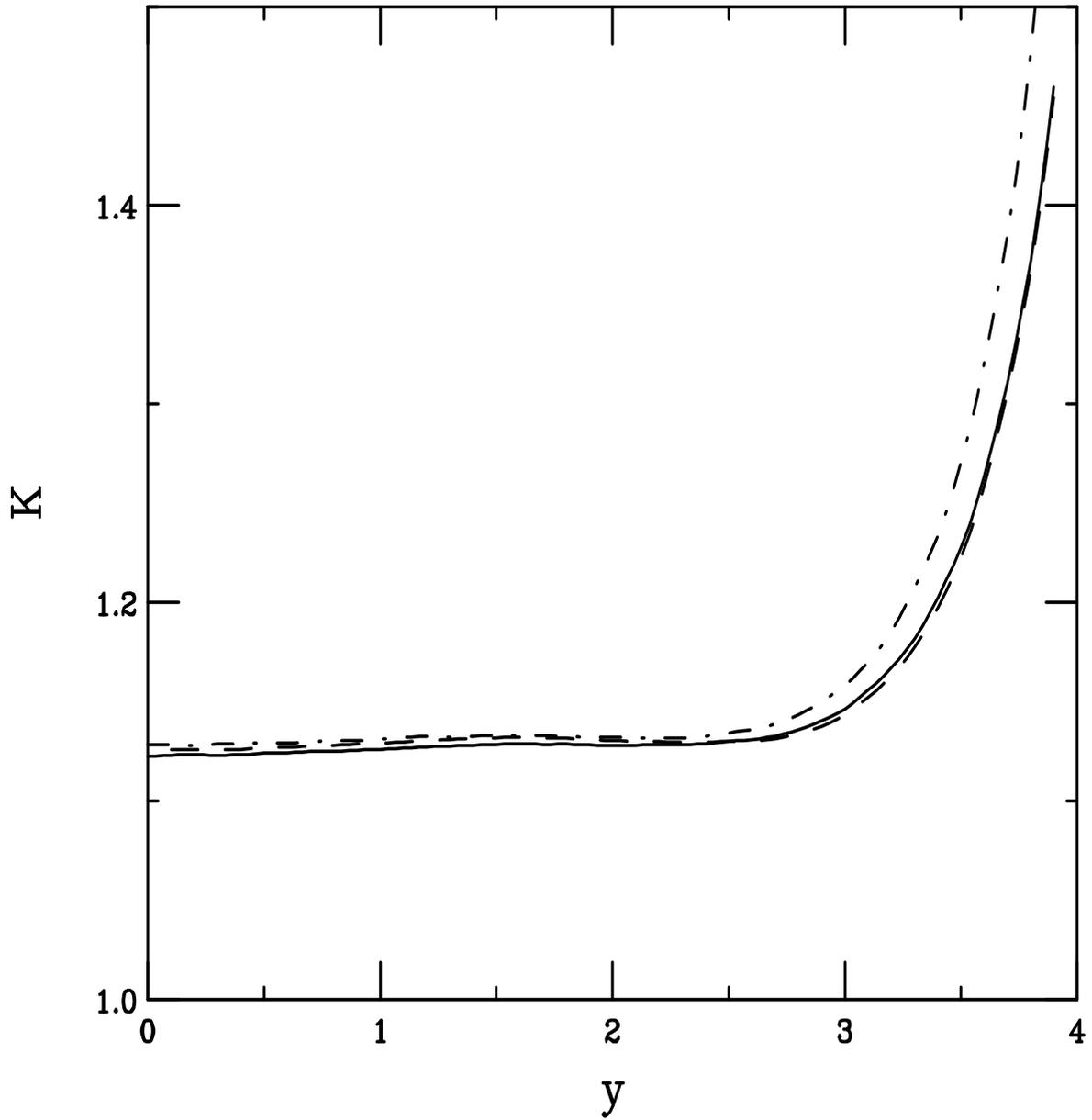}}
\caption[]{The $K$ factors with $S=1$ for $W^+$ (solid), $W^-$ (dashed),
and $Z^0$ (dot-dashed) production are shown.
}
\label{kfac}
\end{figure}

We first show that our shadowing results do not depend strongly on the order
of the calculation.  The ratio of the NLO to LO cross sections, both calculated
with the MRST HO distributions, is often referred to as the
$K$ factor.  The $K$ factor is given as a function of rapidity with no 
shadowing for Pb+Pb collisions at 5.5 TeV in
Fig.~\ref{kfac} for all three vector bosons.  The
$K$ factor is $\approx 1.13$ up to $y=3$.  It grows larger as the edge of phase
space is approached since large, positive, $y$ corresponds to low $x_2$ where
the gluon density is high and the $qg$ channel becomes more important.  The
$qg$ channel
contributes $\approx 15$\% of the total vector boson cross section, with or
without shadowing. The $K$ factor increases faster with $y$
for the $Z^0$ because the higher mass means that 
the phase space for $Z^0$ production is exhausted at lower rapidities than 
$W^\pm$ production, leading to larger $qg$ contributions at high $y_Z$.  
The $K$ factors are quite similar when shadowing is
included and differ from those without shadowing
by $\approx 1$\% at $y=0$.  At high rapidities, a larger
effect might be expected from the gluons but, as seen in Fig.~\ref{shadrat}(c),
the effects of shadowing on the gluon distributions are not as strong as those
of the sea quarks at large $Q^2$.  Thus the difference in the $K$ factors
between the calculations with and without shadowing is only marginally larger
at high rapidity, up to $\approx 4$\% at $y=4$.

\begin{figure}[htb]
\setlength{\epsfxsize=0.95\textwidth}
\setlength{\epsfysize=0.7\textheight}
\centerline{\epsffile{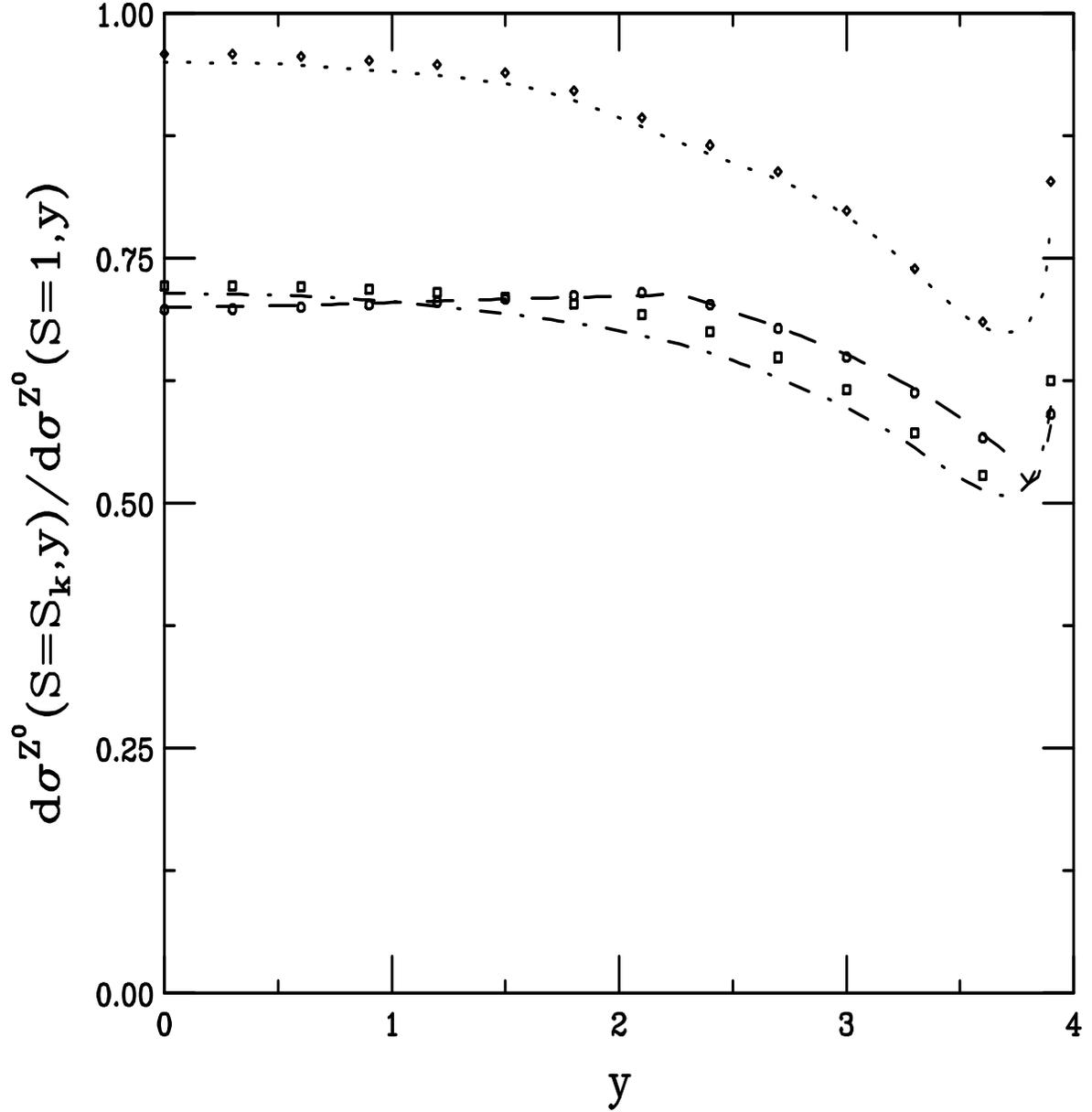}}
\caption[]{The shadowing results at LO and NLO are compared.  The
NLO results are given in the dashed, $S_1$,
dot-dashed, $S_2$, and dotted, $S_3$, lines.  
The LO shadowing ratios for $S_1$, circles, $S_2$,
squares, and $S_3$, diamonds, are also shown.
}
\label{nlovlo}
\end{figure}

In Fig.~\ref{nlovlo}, we show the ratio of $Z^0$ production in Pb+Pb collisions
with and without shadowing at both LO and NLO.  The results are independent of
the order of the calculation, even at large rapidities.  This is not surprising
since we have shown that the differences between shadowing ratios at LO and NLO
are trivial for virtual photon production via the Drell-Yan process
\cite{spenlhc}, even for pairs with $m < m_\Upsilon$.  At the higher scale of
$Z^0$ and $W^\pm$ production, the approximation should be even better because
the $K$ factor is smaller.

We now calculate the NLO $Z^0$, $W^+$
and $W^-$ cross sections in nuclear collisions.
Table~\ref{sigs} gives the total cross sections in the CMS and ALICE central
acceptances, 
$| y | < 2.4$ and $| y | < 1$ respectively, at the LHC.  The cross sections
are larger than the virtual photon mediated
Drell-Yan cross sections at lower masses \cite{spenlhc}.
The results, given for Pb+Pb collisions, are
integrated over impact parameter and presented
in units of nb/nucleon pair.   We note that
with the normalization of $S_{\rm WS}$, the impact-parameter integrated cross
section is unchanged when the spatial dependence is included.  The
next-to-next-to-leading order, NNLO, $W^\pm$ and $Z^0$ total cross sections
have also been calculated \cite{zwrefs}.  The $K$ factors obtained from the
ratio of the ${\cal O}(\alpha \alpha_s^2)$ to ${\cal O}(\alpha)$ cross sections
differs by $\approx 1$\% from the ${\cal O}(\alpha \alpha_s)$ to ${\cal
O}(\alpha)$ $K$ factor shown in Fig.~\ref{kfac}.
Thus changes in the total cross sections between NLO and NNLO
are on the few percent level even though the vector bosons can
be produced in the $gg$ channel as well at NNLO, because
$\alpha_s(m_V^2) \approx 0.116$.   The effects on the shadowing ratios should
be even smaller, see Fig.~\ref{nlovlo}.

We have checked how our results depend on the chosen
set of parton densities.  Using
the CTEQ5M densities \cite{cteq5}, the total cross sections in Table~\ref{sigs}
increase by $\approx 5$\% over the MRST HO results.  However, the $K$ factors
and shadowing ratios in Figs.~\ref{kfac} and \ref{nlovlo} change by less than
1\%.  Thus our shadowing results are essentially independent of parton density.

In Table~\ref{rates} 
we show the expected rate in nucleus-nucleus collisions at $b=0$,
$N(S = S_k) = \sigma_{NN} T_{\rm PbPb}(0) L_{\rm PbPb}^{\rm int} \sigma(S =
S_k)$ with $L_{\rm PbPb}^{\rm int} = 1$/nb in a one month ($10^6$ s) LHC run, 
$\sigma_{NN} = 60$ mb at LHC energies, 
and $T_{\rm PbPb}(0) = 30.4$/mb.  The absolute numbers in 
the experimental acceptances are large but do not reflect the measurable decay
channels.  Including the 3.37\% lepton pair branching for $Z^0$ decays reduces
the number produced with no shadowing, $S=1$, 
to 990 in CMS and 425 in ALICE.  The 10\% lepton 
branching ratio for $W^+$ and $W^-$ leaves nearly 4600 observable decays in
CMS and 1980 in ALICE.

\begin{figure}[h]
\setlength{\epsfxsize=0.95\textwidth}
\setlength{\epsfysize=0.7\textheight}
\centerline{\epsffile{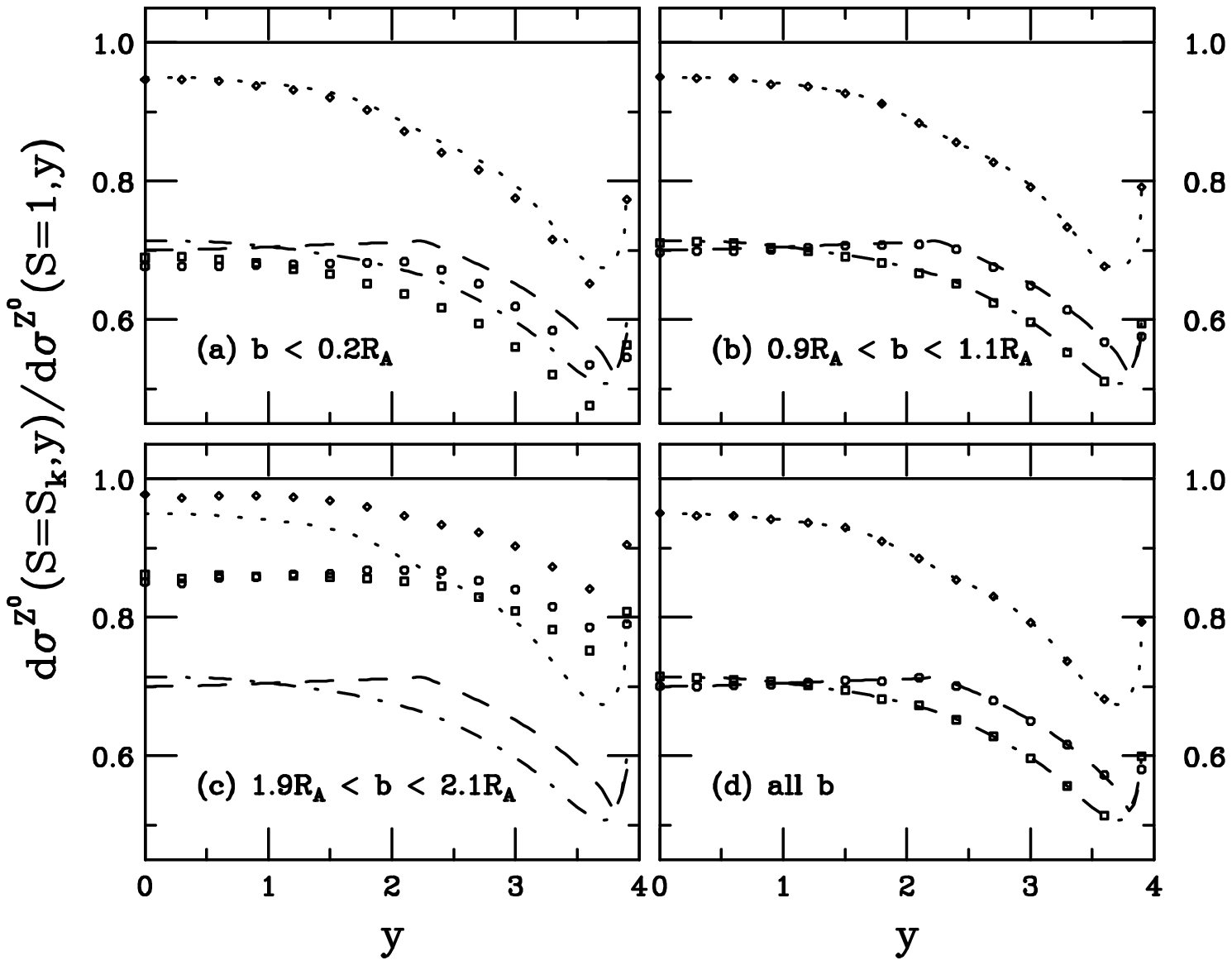}}
\caption{The $Z^0$ rapidity distributions, 
relative to $S=1$ for Pb+Pb collisions at the LHC,
calculated with the MRST HO distributions.  Central, $b<0.2 R_A$,
semi-central, $0.9R_A < b < 1.1R_A$, and peripheral, $1.9R_A < b < 2.1R_A$
impact parameters are shown along with the integral over all $b$.  The
homogeneous shadowing results are given in the dashed, $S_1$,
dot-dashed, $S_2$, and dotted, $S_3$, lines.  
The inhomogeneous
shadowing ratios for $S_{1 \, {\rm WS}}$, circles, $S_{2 \, {\rm WS}}$,
squares, and $S_{3 \, {\rm WS}}$, diamonds, are also shown.
}
\label{zrap}
\end{figure}

\begin{figure}[h]
\setlength{\epsfxsize=0.95\textwidth}
\setlength{\epsfysize=0.7\textheight}
\centerline{\epsffile{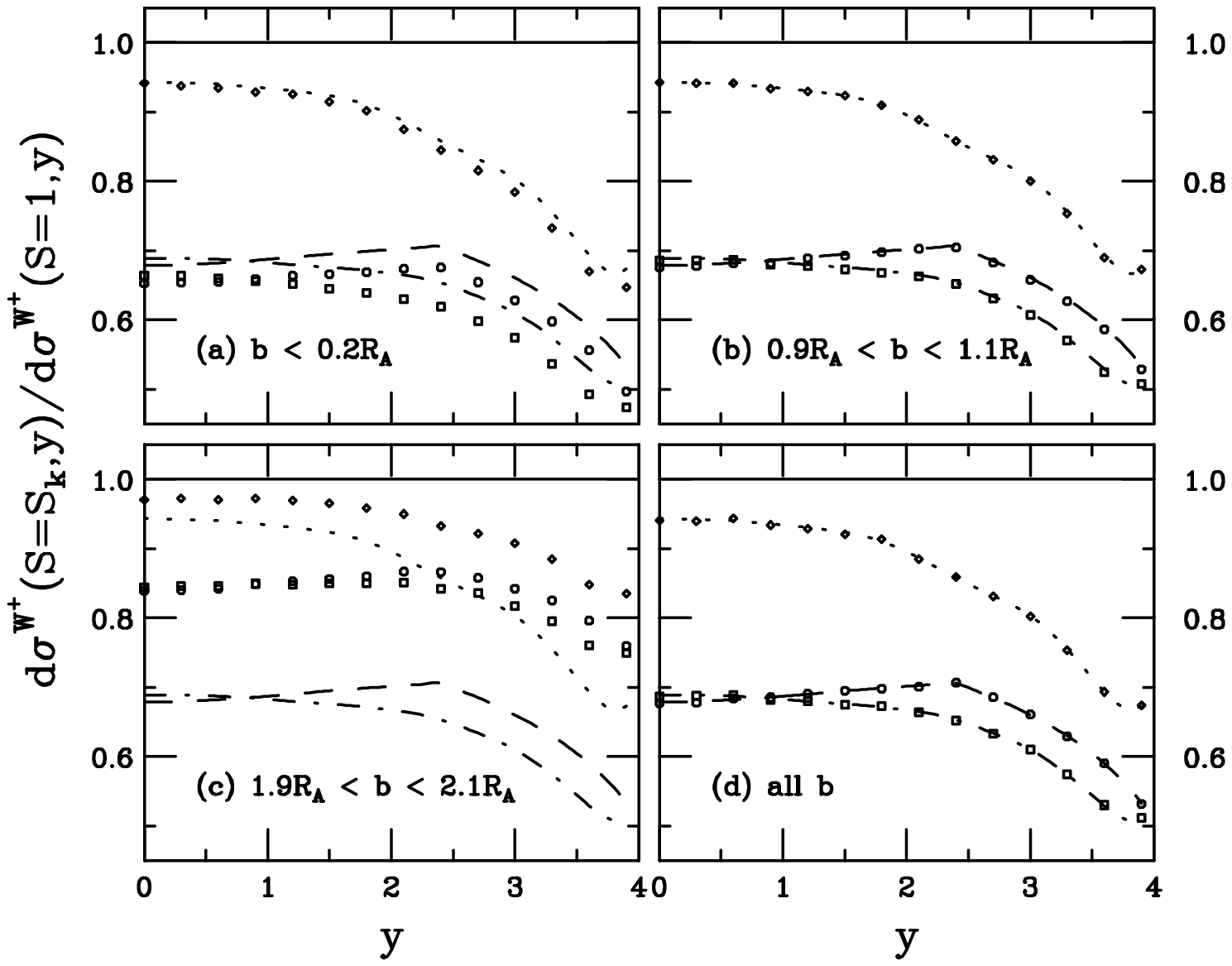}}
\caption{The $W^+$ rapidity distributions, 
relative to $S=1$ for Pb+Pb collisions at the LHC,
calculated with the MRST HO distributions.  Central, $b<0.2 R_A$,
semi-central, $0.9R_A < b < 1.1R_A$, and peripheral, $1.9R_A < b < 2.1R_A$
impact parameters are shown along with the integral over all $b$.  The
homogeneous shadowing results are given in the dashed, $S_1$,
dot-dashed, $S_2$, and dotted, $S_3$, lines.  
The inhomogeneous
shadowing ratios for $S_{1 \, {\rm WS}}$, circles, $S_{2 \, {\rm WS}}$,
squares, and $S_{3 \, {\rm WS}}$, diamonds, are also shown.
}
\label{wprap}
\end{figure}

Figures \ref{zrap} and \ref{wprap} compare the ratios of $Z^0$ and $W^+$ 
production in Pb+Pb collisions with the three shadowing parameterizations to
Pb+Pb collisions with no shadowing as a function of rapidity.  The isospin
effects wash out the differences between the $W^+$ and $W^-$ distributions in
the ratios so that the results are essentially identical for the two charged
vector bosons.  Therefore the ratios are shown only for the $W^+$.
The results are given for several impact
parameter bins, the most central bin, $b < 0.2 R_A$, an intermediate impact
parameter bin around $b \sim R_A$, and a peripheral bin around $b \sim 2R_A$.
It is clear that by neglecting the impact parameter dependence of shadowing,
one may overestimate the effect in peripheral collisions, an 
important point if using the $Z^0$ as a baseline in different transverse energy
bins.  The integration over all impact parameters is equivalent
to the average shadowing, as expected from the normalization of 
Eq.~(\ref{wsparam}).  Although the results are shown using the local density
approximation, the parameterization of the spatial dependence for a long 
coherence length, Eq.~(\ref{rhoparam}), differs only marginally.  In central
collisions, the difference between the two parameterizations is less than 1\%
while in the most peripheral bin, it is $3-6$\%.  The largest differences occur
in regions with the strongest shadowing modifications.
Thus the calculations are rather
insensitive to the exact spatial parameterization, suggesting that heavy ion
collisions cannot distinguish between different dependencies, only between
homogeneous and inhomogeneous shadowing.

The ratios are rather similar for all vector bosons.  The $S_1$ and $S_2$
ratios are approximately equal as a function of rapidity, presumably because
the $Q^2$ evolution of the $S_2$ parameterization ends at $Q^2 = 100$ GeV$^2$. 
The calculations cover the entire rapidity range of vector boson
production.  At $y_Z \sim 0$, $x_1 = x_2 = 0.017$, in the low $x$ region.
As rapidity increases, $x_1$ increases, going through the antishadowing region
and the EMC region with $x_1 \sim 0.33$ at $y_Z = 3$.  The kink in the $S_1$
ratio at $y\sim 2.2$ is an artifact resulting from the rather sharp transition
between the shadowing and EMC regions at $x \sim 0.15$, see Fig.~\ref{shadrat}.
Note that the larger
rapidity coverage of CMS makes the EMC region accessible in this measurement.
When $y_Z \rightarrow
4$, $x_1 \rightarrow 1$, entering the ``Fermi motion'' region and causing the
upturn of the ratios at large $y_Z$.  Note also that at large $x_1$, 
the valence quarks dominate.  While increasing $y_Z$ ($x_1$) traces out the 
large $x$ portion of the shadowing curve, the low $x$ part of the 
shadowing regime is accessible from $x_2$ with growing $y_Z$.  At $y_Z = 3$,
$x_2 \sim 8 \times 10^{-4}$, in a range where shadowing saturates in $S_1$ and
$S_2$.  There is no saturation built into the $S_3$ parameterization, causing
a steeper decrease in the ratios for large $y_Z$ with this parameterization
than with $S_1$ and $S_2$.  In addition, the $S_3$ sea quark shadowing is 
never as strong at low $x$ as for $S_1$ and $S_2$ so that these two 
parameterizations are both more strongly shadowed overall.  
The $Z^0$ ratios are all slightly higher than those for $W^\pm$ 
because the larger mass of the $Z^0$ results in $x_Z \sim 1.1 x_{W}$.

We also point out that the large vector boson masses do not allow us to
restrict ourselves only to an $x$ region where the coherence length is always
larger than $R_A$
so that the multiple interaction approach could be used without having 
to match the spatial dependence across $x$ boundaries.  While the target 
parton is at relatively low $x$, the projectile parton, also affected by 
shadowing, is at relatively high $x$ where $l_c$ is small.

\begin{figure}[htb]
\setlength{\epsfxsize=0.95\textwidth}
\setlength{\epsfysize=0.7\textheight}
\centerline{\epsffile{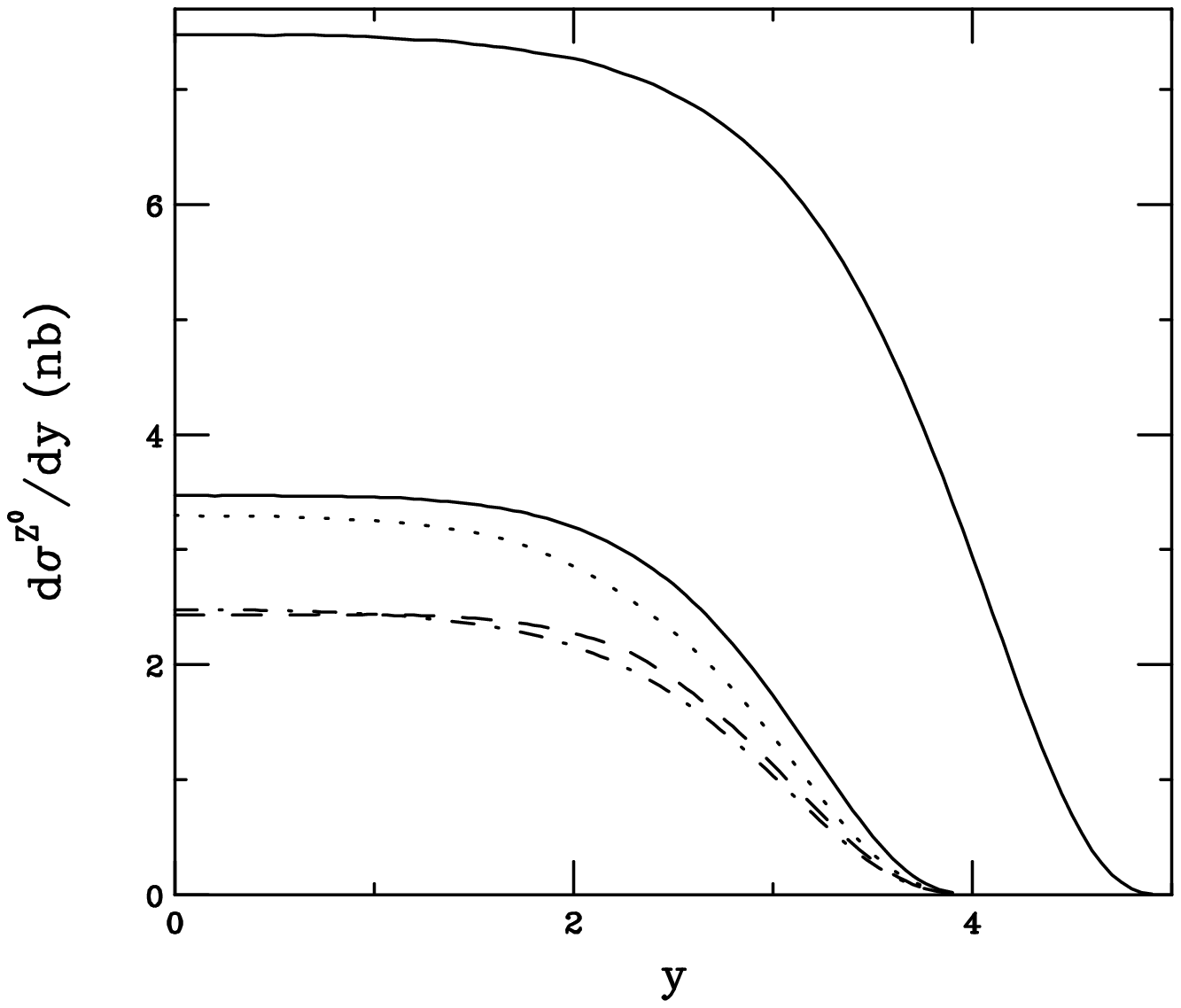}}
\caption[]{The $Z^0$ rapidity distributions in $pp$ and Pb+Pb collisions,
calculated with the MRST HO distributions.  The upper solid curve is
the $pp$ result at 14 TeV while the lower solid curve is the Pb+Pb
distribution at 5.5 TeV/nucleon pair with no shadowing.
The homogeneous shadowing results for Pb+Pb collisions 
are given in the dashed, $S_1$, dot-dashed, $S_2$, and dotted, $S_3$, lines.  
}
\label{zdists}
\end{figure}

\begin{figure}[htb]
\setlength{\epsfxsize=0.95\textwidth}
\setlength{\epsfysize=0.7\textheight}
\centerline{\epsffile{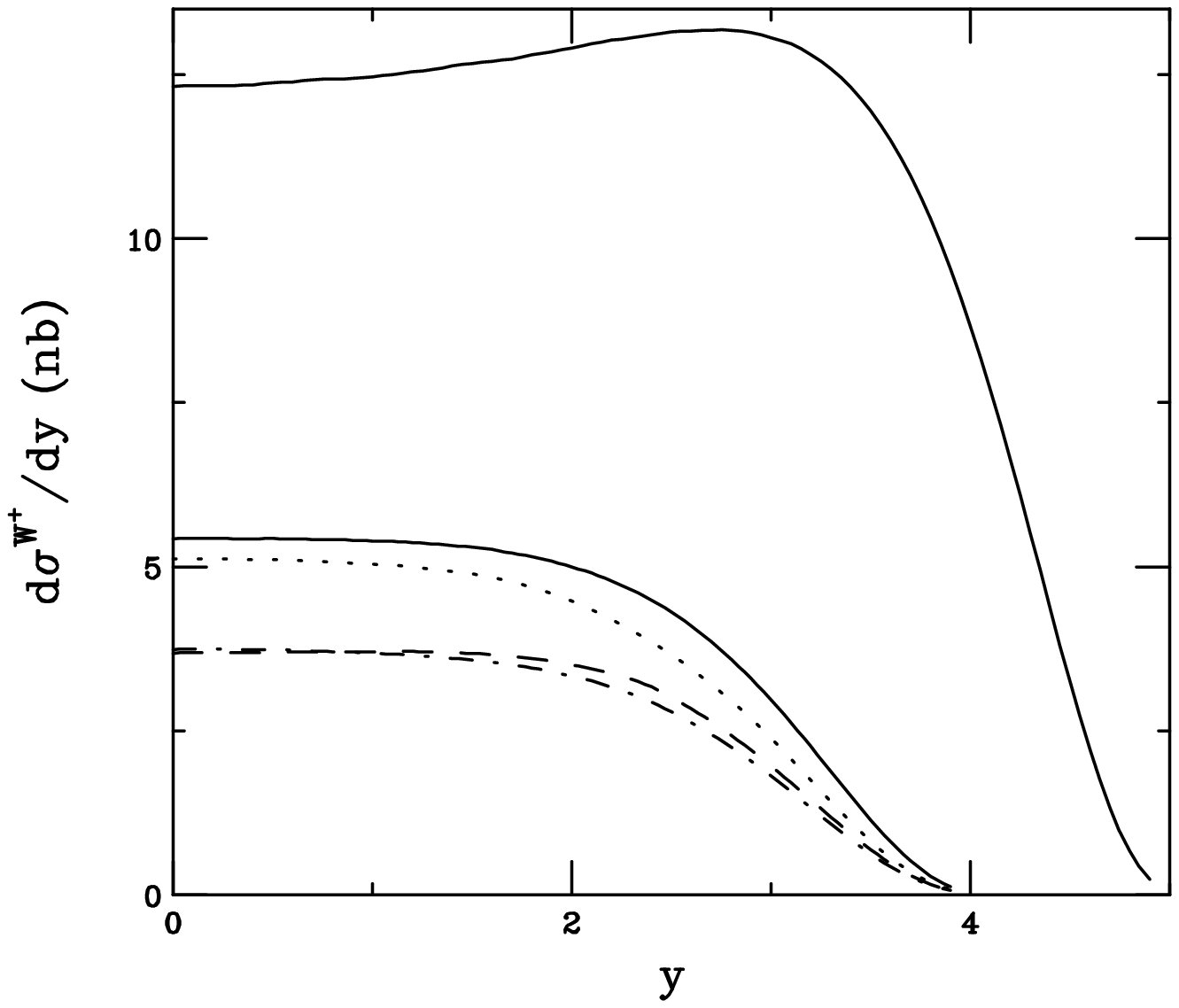}}
\caption[]{The $W^+$ rapidity distributions in $pp$ and Pb+Pb collisions,
calculated with the MRST HO distributions.  The upper solid curve is
the $pp$ result at 14 TeV while the lower solid curve is the Pb+Pb
distribution at 5.5 TeV/nucleon pair with no shadowing.
The homogeneous shadowing results for Pb+Pb collisions 
are given in the dashed, $S_1$, dot-dashed, $S_2$, and dotted, $S_3$, lines.  
}
\label{wpdists}
\end{figure}

\begin{figure}[htb]
\setlength{\epsfxsize=0.95\textwidth}
\setlength{\epsfysize=0.7\textheight}
\centerline{\epsffile{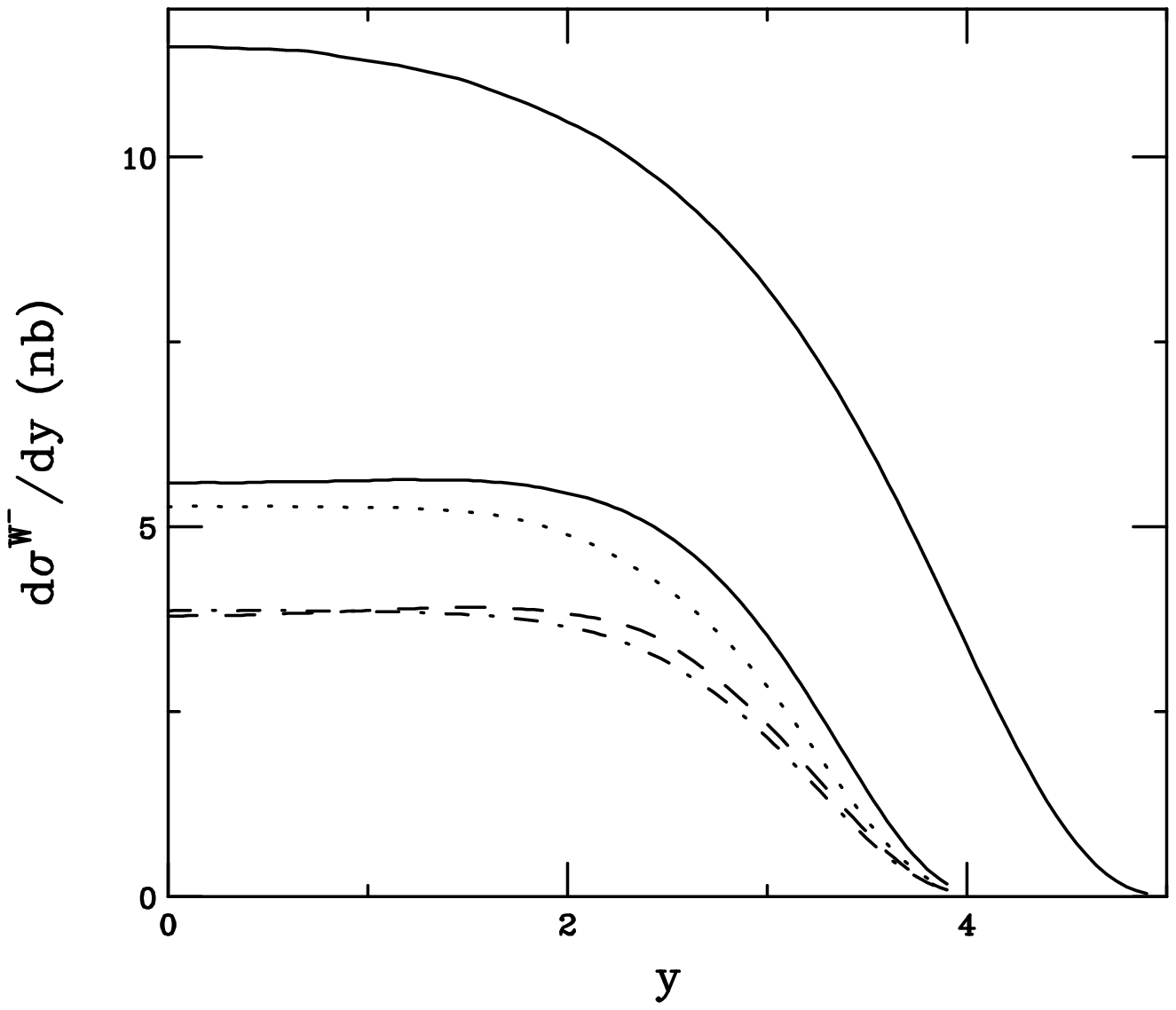}}
\caption[]{The $W^-$ rapidity distributions in $pp$ and Pb+Pb collisions,
calculated with the MRST HO distributions.  The upper solid curve is
the $pp$ result at 14 TeV while the lower solid curve is the Pb+Pb
distribution at 5.5 TeV/nucleon pair with no shadowing.
The homogeneous shadowing results for Pb+Pb collisions 
are given in the dashed, $S_1$, dot-dashed, $S_2$, and dotted, $S_3$, lines.  
}
\label{wmdists}
\end{figure}

The shadowing ratios are fairly simply traced out for vector boson production,
especially at leading order since the fixed boson mass defines $x$ at any
$y$ whereas Drell-Yan shadowing effects are smeared over
the mass interval.  However, the 
ratios shown in Figs.~\ref{zrap} and \ref{wprap} will not be accessible 
experimentally due to the nuclear isospin.  The comparison must be made to $pp$
interactions, preferably at the same energy to retain the same $x$ values.
This ideal situation may not be realized for some time at the LHC.  Therefore
in Figs.~\ref{zdists}-\ref{wmdists} we show the Pb+Pb rapidity distributions
with and without homogeneous
shadowing as well as the distributions from $pp$ collisions
at 14 TeV for all three vector bosons.  The Pb+Pb cross
sections are given per nucleon pair for a more
direct comparison.  The higher energy
extends the available vector boson rapidity space by one unit.  The $Z^0$
distributions in Fig.~\ref{zdists} have a plateau over two units of rapidity.  
The $pp$ $W^+$ distribution in Fig.~\ref{wpdists}
rises over the first several units
of rapidity, followed by a decrease as the edge of phase space is approached.
This is due to the increasing importance of valence quarks
at large $y$ (large $x_1$).  The effect appears for $W^+$ production in $pp$ 
collisions because the $u$ valence quarks carry more momentum than the $d$
valence quarks, see Fig.~\ref{pdfz} (a).  On the other hand, the $W^-$ $pp$
distribution always decreases with rapidity.  In Pb+Pb collisions, there is a
slight increase in $W^-$ production with rapidity instead of a decrease while
the $W^+$ distributions are either flat or decreasing.  This increase
in $W^-$ production shown in Fig.~\ref{wmdists}
is due to the neutron excess in Pb+Pb where, in
$nn$ collisions, $W^-$ production proceeds dominantly through 
$f_{\overline u}^n f_d^n (\approx f_{\overline d}^p f_u^p)$.  Likewise, the
rise in $W^+$ production in Pb+Pb relative to $pp$ collisions
disappears because of the neutron content of the nucleus.

\begin{figure}[htb]
\setlength{\epsfxsize=0.95\textwidth}
\setlength{\epsfysize=0.7\textheight}
\centerline{\epsffile{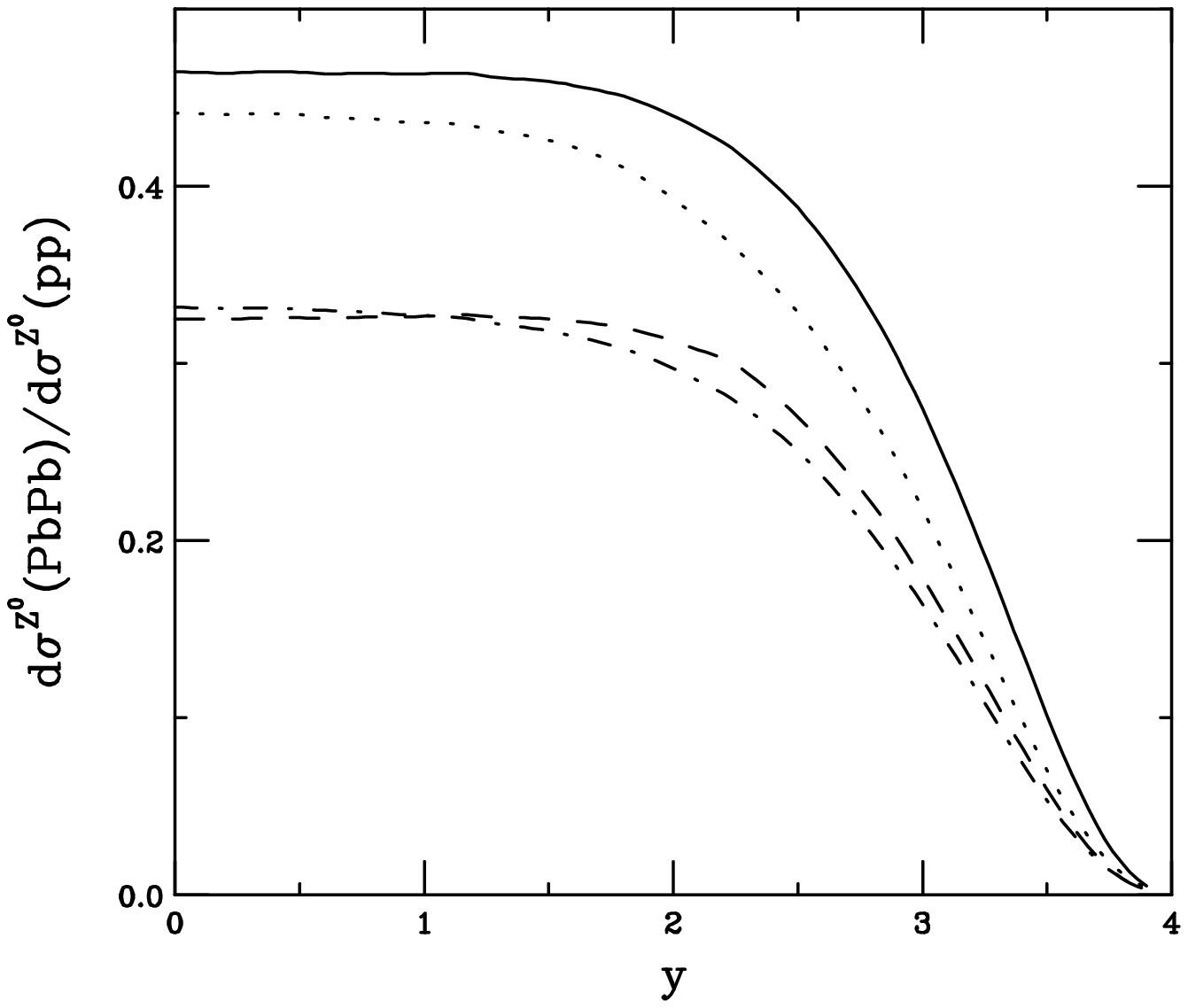}}
\caption[]{The ratios of the $Z^0$ rapidity distributions in Pb+Pb 
collisions relative to $pp$ collisions,
calculated with the MRST HO distributions.  The solid curve is the ratio
without shadowing.
The homogeneous shadowing results are given in the dashed, $S_1$,
dot-dashed, $S_2$, and dotted, $S_3$, lines.  
}
\label{zpb2pp}
\end{figure}

\begin{figure}[htb]
\setlength{\epsfxsize=0.95\textwidth}
\setlength{\epsfysize=0.7\textheight}
\centerline{\epsffile{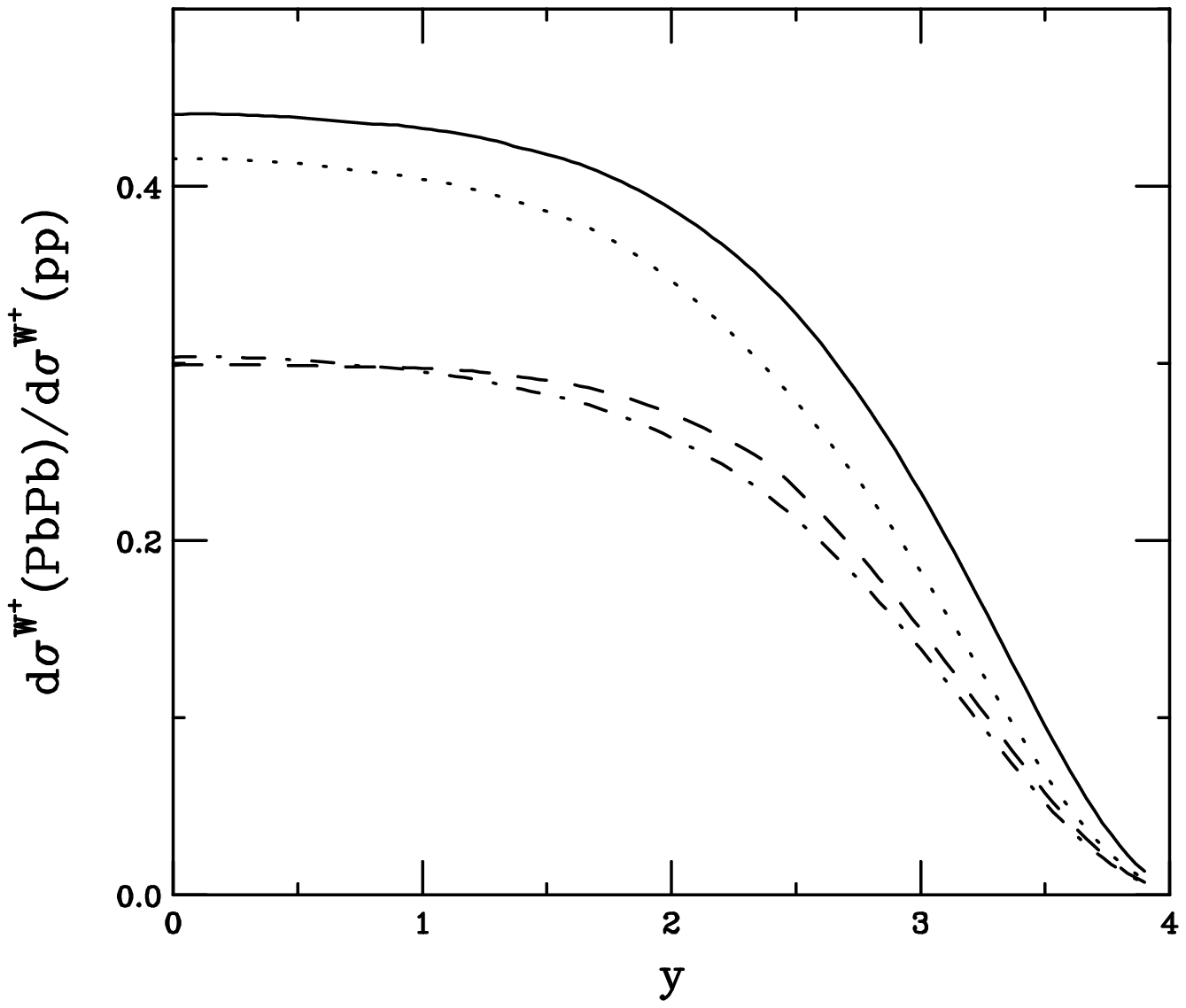}}
\caption[]{The ratios of the $W^+$ rapidity distributions in Pb+Pb 
collisions relative to $pp$ collisions,
calculated with the MRST HO distributions.  The solid curve is the ratio
without shadowing.
The homogeneous shadowing results are given in the dashed, $S_1$,
dot-dashed, $S_2$, and dotted, $S_3$, lines.  
}
\label{wppb2pp}
\end{figure}

\begin{figure}[htb]
\setlength{\epsfxsize=0.95\textwidth}
\setlength{\epsfysize=0.7\textheight}
\centerline{\epsffile{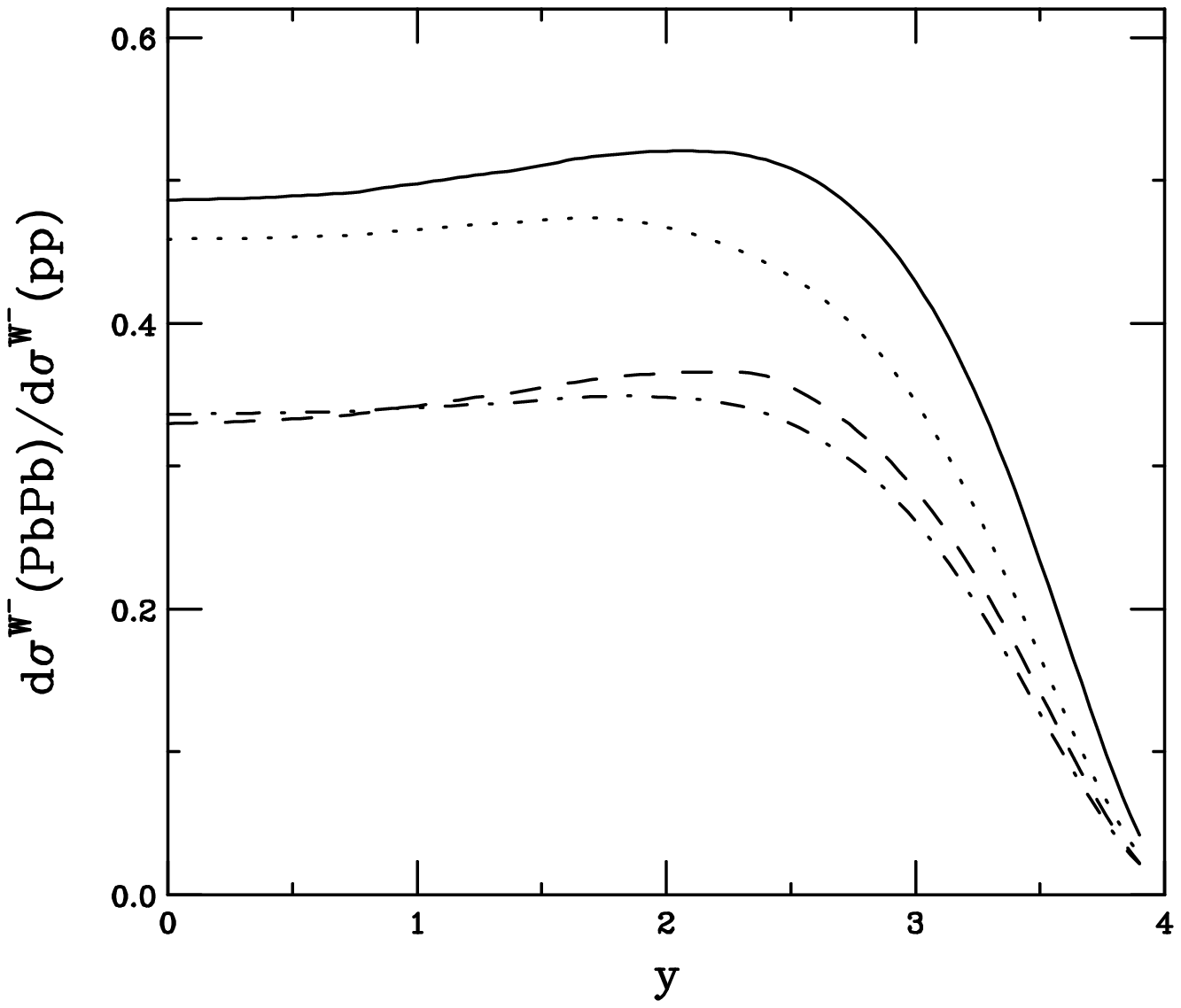}}
\caption[]{The ratios of the $W^-$ rapidity distributions in Pb+Pb 
collisions relative to $pp$ collisions,
calculated with the MRST HO distributions.  The solid curve is the ratio
without shadowing.
The homogeneous shadowing results are given in the dashed, $S_1$,
dot-dashed, $S_2$, and dotted, $S_3$, lines.  
}
\label{wmpb2pp}
\end{figure}

Finally, the ratios of the Pb+Pb/nucleon pair to $pp$ cross sections are shown 
in Figs.~\ref{zpb2pp}-\ref{wmpb2pp} for homogeneous shadowing.  Due to the 
higher $pp$ cross sections, the ratios are lower than those to $S=1$ at 5.5 
TeV shown in Figs.~\ref{zrap} and \ref{wprap}.  
Since both the Pb+Pb and $pp$ $Z^0$ rapidity distributions are rather flat,
the ratios in Fig.~\ref{zpb2pp} are also flat to $y\sim 1.5$.  The rise in the
$W^+$ $pp$ distribution shown in Fig.~\ref{wpdists} causes the Pb+Pb to $pp$
ratios to decrease with rapidity over all $y$ in Fig.~\ref{wppb2pp}.  
However, the increase in $W^-$
production due to neutrons in Pb+Pb collisions, barely visible in
Fig.~\ref{wmdists}, is apparent in the $W^-$ Pb+Pb to $pp$ ratio in
Fig.~\ref{wmpb2pp}. 
It should still be possible to distinguish between the shadowing
parameterizations and study quark shadowing at $Q^2=m_V^2$, 
particularly since the 14 TeV $pp$ data will be available with higher
statistics.  Note that comparing the $pp$ results to the Pb+Pb
calculations with inhomogeneous shadowing would result in slightly lower ratios
in central collisions and higher ratios in peripheral collisions, as expected
from Figs.~\ref{zrap} and \ref{wprap}.

Once the basic nuclear shadowing effects on vector boson production have been
understood, they can perhaps be used to study other medium effects in heavy
ion collisions by comparing the leptonic and hadronic decay channels.
The hadronic decays of the vector bosons, $\sim 70$\% of all decays of each
boson, may be more difficult to interpret.  While the width of the $Z^0$
decay to $l^+ l^-$ is not expected to be modified in the quark-gluon plasma
due to the weak coupling
\cite{kapustawong}, the $Z^0$ has a 2.49 GeV total width and will decay in any 
quark-gluon plasma to two jets through $Z^0 \rightarrow q \overline q 
\rightarrow {\rm jet} \, + \, {\rm jet}$ in $\sim 0.1$ fm.  Therefore, the
decay jets could be modified in the medium 
which may still be progressing toward
thermalization and will be subject to rescattering and jet quenching.  Thus
a comparison of a reconstructed $Z^0$ in the dilepton channel where no nuclear
effects are expected and medium-modified
jets should result in a broader width in the $q \overline q$ channel than the
$l^+ l^-$ channel \cite{Geist}.  In addition, the $Z^0$ could be used
to tag jets through the $q \overline q \rightarrow Z^0 g$ and $gq
\rightarrow
Z^0 q$ channels to study jet properties in the quark-gluon plasma
\cite{kvatz}.

{\bf Acknowledgements} I thank K.J. Eskola for providing the shadowing 
parameterizations.  I thank D. Kharzeev, K. Redlich and 
U.A. Wiedemann for discussions.

\setcounter{equation}{0}
\renewcommand{\theequation}{A\arabic{equation}}
\begin{center}
{\bf Appendix}
\end{center}
\vspace{0.2in}

In $AB$ collisions, the cross section per nucleon
must include the nuclear isospin since, in general, $\sigma_{pp}^V \neq
\sigma_{pn}^V \neq \sigma_{np}^V \neq \sigma_{nn}^V$.
We give the convolution of the nuclear
parton densities in Eq.~(\ref{sigmajpsi}), 
including only the couplings.  We take $f_{d_V}^n = f_{u_V}^p$, $f_{u_V}^n = 
f_{d_V}^p$, $f_{\overline d}^n = f_{\overline u}^p$, and $f_{\overline u}^n = 
f_{\overline d}^p$.  All other distributions are assumed to be identical for 
protons and neutrons. The proton and neutron numbers in nucleus $A$ are $Z_A$ 
and $N_A$.  To be concise, we define
\begin{eqnarray}
S^u(A,x)f_u^p(x,Q^2) & = & S^{u_V}(A,x) f_{u_V}^p + S^{\overline u}(A,x)
f_{\overline u}^p \\
S^d(A,x)f_d^p(x,Q^2) & = & S^{d_V}(A,x) f_{d_V}^p + S^{\overline d}(A,x)
f_{\overline d}^p \, \, 
\end{eqnarray}
where we have abbreviated the shadowing functions as $S^i(A,x)$.

We begin with the $q \overline q$ channel.  For $Z^0$ production, we have
\begin{eqnarray}
\lefteqn{ \sum_{i,j \in Q \overline Q} 
S^i(A,x_1) S^j(B,x_2)f_{q_i}^N(x_1,Q^2) f_{\overline 
q_j}^N(x_2,Q^2) C^{\rm ii}(q_i, \overline q_j) [(g_V^i)^2 
+ (g_A^i)^2]}
\label{dszdyqqb} \\ & & \mbox{}  =
\frac{1}{8}[1 - \frac{8}{3}x_W + \frac{32}{9}x_W^2] 
\left( S^u(A,x_1) S^{\overline u}(B,x_2)
\left\{ Z_A f_u^p(x_1,Q^2)+ N_A f_u^n(x_1,Q^2) \right\} \right.
\nonumber \\ & & \mbox{} \times
\left. \left\{ Z_B f_{\overline u}^p(x_2,Q^2) + N_B 
f_{\overline u}^n(x_2,Q^2)\right\} + 2 AB S^c(A,x_1) S^{\overline c}(B,x_2)
f_c^p(x_1,Q^2) f_{\overline c}^p(x_2,Q^2)
\right)
\nonumber \\ & & \mbox{} +
\frac{1}{8}[1 - \frac{4}{3}x_W + \frac{8}{9}x_W^2] 
\left( S^d(A,x_1) S^{\overline d}(B,x_2)
\left\{ Z_A f_d^p(x_1,Q^2)+ N_A f_d^n(x_1,Q^2) \right\} \right. 
\nonumber \\ & & \mbox{} \times \left. \left.\left.
\left\{ Z_B f_{\overline d}^p(x_2,Q^2) + N_B 
f_{\overline d}^n(x_2,Q^2) \right\} +
2 AB S^s(A,x_1) S^{\overline s}(B,x_2)
f_s^p(x_1,Q^2) f_{\overline s}^p(x_2,Q^2) \right) \right. \right. \nonumber \\ 
& & \mbox{} + [ x_1 \leftrightarrow x_2 , A \leftrightarrow B ] \, \, .
\nonumber
\end{eqnarray}
Note that for $Z^0$ production, we have also included the square of the vector
and axial vector couplings since these depend on the quark charges.
The $W^+$ $q \overline q$ convolution is
\begin{eqnarray}
\lefteqn{ \sum_{i,j \in Q, \overline Q} 
S^i(A,x_1) S^j(B,x_2)f_{q_i}^N(x_1,Q^2) f_{\overline 
q_j}^N(x_2,Q^2) C^{\rm ii}(q_i, \overline q_j)}
\label{dswpdyqqb} \\ & & \mbox{} =
\cos^2 \theta_C
\left( S^u(A,x_1) S^{\overline d}(B,x_2)
\left\{ Z_A f_u^p(x_1,Q^2)+ N_A f_u^n(x_1,Q^2) \right\} \right.
\nonumber \\ & & \mbox{} \times \left.
\left\{ Z_B f_{\overline d}^p(x_2,Q^2) + N_B 
f_{\overline d}^n(x_2,Q^2)\right\} + ABS^{\overline s}(A,x_1) S^c(B,x_2)
f_{\overline s}^p(x_1,Q^2) f_c^p(x_2,Q^2) \right)
\nonumber \\ & & \mbox{} + \left. \sin^2 \theta_C
\left( S^u(A,x_1)S^{\overline s}(B,x_2)
\left\{ Z_A f_u^p(x_1,Q^2)+ N_A f_u^n(x_1,Q^2) \right\} B
  f_{\overline s}^p (x_2,Q^2) \right. \right. \nonumber \\ & & \mbox{} 
\left. \left. + S^{\overline d}(A,x_1)S^c(B,x_2)
\left\{ Z_A f_{\overline d}^p(x_1,Q^2)+ N_A f_{\overline d}^n(x_1,Q^2) 
\right\} B f_c^p (x_2,Q^2) 
\right) \right\} \nonumber \\ & & \mbox{}
+ [ x_1 \leftrightarrow x_2, A \leftrightarrow B 
] \, \, . \nonumber
\end{eqnarray}
Finally, the $W^-$ $q \overline q$ convolution is
\begin{eqnarray}
\lefteqn{ \sum_{i,j\in Q \overline Q} 
S^i(A,x_1) S^j(B,x_2)f_{q_i}^N(x_1,Q^2) f_{\overline q_j}^N(x_2,Q^2) 
C^{\rm ii}(q_i, \overline q_j)}
\label{dswmdyqqb} \\ & & \mbox{}  =
\cos^2 \theta_C
\left( S^{\overline u}(A,x_1) S^d(B,x_2)
\left\{ Z_A f_{\overline u}^p(x_1,Q^2)+ N_A f_{\overline
u}^n(x_1,Q^2) \right\} \right.
\nonumber \\ & & \mbox{} \times \left.
\left\{ Z_B f_d^p(x_2,Q^2) + N_B 
f_d^n(x_2,Q^2)\right\} + ABS^s(A,x_1) S^{\overline c}(B,x_2)
f_s^p(x_1,Q^2) f_{\overline c}^p(x_2,Q^2) \right)
\nonumber \\ & & \mbox{} + \left. \sin^2 \theta_C
\left( S^{\overline u}(A,x_1) S^s(B,x_2)
\left\{ Z_A f_{\overline u}^p(x_1,Q^2)+ 
N_A f_{\overline u}^n(x_1,Q^2) \right\} B
  f_s^p (x_2,Q^2) \right. \right. \nonumber \\ & & \mbox{} 
\left. \left. + S^d(A,x_1)S^{\overline c}(B,x_2)
\left\{ Z_A f_d^p(x_1,Q^2)+ N_A f_d^n(x_1,Q^2) 
\right\} B f_{\overline c}^p (x_2,Q^2) 
\right) \right\}  \nonumber \\ & & \mbox{}
+ [ x_1 \leftrightarrow x_2 , A \leftrightarrow B ] \, \, . \nonumber
\end{eqnarray}

We now turn to the $qg$ channel. The convolution for $Z^0$ production is
\begin{eqnarray}
\lefteqn{ \sum_{i,k \in Q \overline Q} \bigg(
S^i(A,x_1) S^g(B,x_2)f_{q_i}^N(x_1,Q^2) f_g^N(x_2,Q^2) } \nonumber \\
&  & \mbox{} + 
[ x_1 \leftrightarrow x_2, A \leftrightarrow B ] \bigg)
C^{\rm if}(q_i, q_k) [(g_V^i)^2 
+ (g_A^i)^2]
\label{dszdyqg} \\ & & \mbox{}  = BS^g(B,x_2)f_g^p(x_2,Q^2) \bigg\{
\frac{1}{8}[1 - \frac{8}{3}x_W + \frac{32}{9}x_W^2] 
\left( S^u(A,x_1) 
\left\{ Z_A f_u^p(x_1,Q^2)+ N_A f_u^n(x_1,Q^2) \right\} \right. 
\nonumber \\ & & \mbox{} \left. + S^{\overline u}(A,x_1) \left\{ Z_B 
f_{\overline u}^p(x_2,Q^2) + N_B 
f_{\overline u}^n(x_2,Q^2)\right\} + 2 AS^c(A,x_1)f_c^p(x_1,Q^2) 
\right)
\nonumber \\ & & \mbox{} +
\frac{1}{8}[1 - \frac{4}{3}x_W + \frac{8}{9}x_W^2] 
\left( S^d(A,x_1) 
\left\{ Z_A f_d^p(x_1,Q^2)+ N_A f_d^n(x_1,Q^2) \right\} \right. 
\nonumber \\ & & \mbox{} + \left. \left.\left. S^{\overline d}(A,x_1)
\left\{ Z_B f_{\overline d}^p(x_2,Q^2) + N_B 
f_{\overline d}^n(x_2,Q^2) \right\} +
2 A S^s(A,x_1) f_s^p(x_1,Q^2) \right) \right. \right. \bigg\} 
\nonumber \\ 
& & \mbox{} + [ x_1 \leftrightarrow x_2, A \leftrightarrow B ] \, \, .
\nonumber
\end{eqnarray}
For $W^+$ production in the $qg$ channel, we have
\begin{eqnarray}
\lefteqn{ \sum_{i,k \in Q \overline Q} \bigg(
S^i(A,x_1) S^g(B,x_2)f_{q_i}^N(x_1,Q^2) f_g^N(x_2,Q^2) + 
[ x_1 \leftrightarrow x_2, A \leftrightarrow B ] \bigg)
C^{\rm if}(q_i, q_k)}
\label{dswpdyqg} \\ & & \mbox{}  = BS^g(B,x_2)f_g^p(x_2,Q^2) \bigg[
S^u(A,x_1) 
\left\{ Z_A f_u^p(x_1,Q^2)+ N_A f_u^n(x_1,Q^2) \right\}  
\nonumber \\ & & \mbox{} \left. + S^{\overline d}(A,x_1) \left\{ Z_B 
f_{\overline d}^p(x_2,Q^2) + N_B 
f_{\overline d}^n(x_2,Q^2)\right\} 
+ A \left\{S^{\overline s}(A,x_1)f_{\overline
s}^p (x_1,Q^2) + S^c(A,x_1)f_c^p(x_1,Q^2) \right\}  \bigg] \right.
\nonumber \\ 
& & \mbox{} + [ x_1 \leftrightarrow x_2, A \leftrightarrow B ] \, \, .
\nonumber
\end{eqnarray}
Now the couplings do not enter explicitly for $W^+$ and $W^-$ production 
because each
distribution is multiplied by $(\cos^2 \theta_C + \sin^2 \theta_C)$.  Finally,
the $qg$ convolution for $W^-$ production is 
\begin{eqnarray}
\lefteqn{ \sum_{i,k \in Q \overline Q} \bigg(
S^i(A,x_1) S^g(B,x_2)f_{q_i}^N(x_1,Q^2) f_g^N(x_2,Q^2) + 
[ x_1 \leftrightarrow x_2, A \leftrightarrow B ] \bigg)
C^{\rm if}(q_i, q_k)}
\label{dswmdyqg} \\ & & \mbox{}  = BS^g(B,x_2)f_g^p(x_2,Q^2) \bigg[
S^{\overline u}(A,x_1) 
\left\{ Z_A f_{\overline u}^p(x_1,Q^2)+ N_A f_{\overline u}^n(x_1,Q^2) 
\right\}  
\nonumber \\ & & \mbox{} \left. + S^d(A,x_1) \left\{ Z_B 
f_d^p(x_2,Q^2) + N_B f_d^n(x_2,Q^2)\right\} 
+ A \left\{S^s(A,x_1)f_s^p (x_1,Q^2) + S^{\overline c}(A,x_1)f_{\overline
 c}^p(x_1,Q^2) \right\}  \bigg] \right.
\nonumber \\ 
& & \mbox{} + [ x_1 \leftrightarrow x_2, A \leftrightarrow B ] \, \, .
\nonumber
\end{eqnarray}

\begin{table}[htb]
\begin{tabular}{ccccc}
Detector & $\sigma(S=1)$ (nb) &  $\sigma(S=S_1)$ (nb) &
$\sigma(S=S_2)$ (nb) & $\sigma(S=S_3)$ (nb) \\ \hline
\multicolumn{5}{c}{$Z^0$} \\
CMS    & 16.10 & 11.37 & 11.22 & 14.92 \\
ALICE  &  6.93 &  4.87 &  4.93 &  6.56 \\
\multicolumn{5}{c}{$W^+$} \\
CMS    & 25.18 & 17.39 & 17.08 & 23.23 \\
ALICE  & 10.84 &  7.39 &  7.45 & 10.19 \\
\multicolumn{5}{c}{$W^-$} \\
CMS    & 26.63 & 18.39 & 18.12 & 24.58 \\
ALICE  & 11.21 &  7.64 &  7.73 & 10.54 \\
\end{tabular}
\caption{Vector boson production cross sections in units of nb per
nucleon pair in Pb+Pb collisions at 5.5 TeV/nucleon
calculated with the MRST HO parton densities. 
Full azimuthal coverage is assumed.  The corresponding $pp$ cross
sections at 14 TeV are $\sigma^{Z^0} = 35.44$ nb (CMS), 14.94 nb
(ALICE), $\sigma^{W^+} = 60.50$ nb (CMS), 24.76 nb (ALICE) and
$\sigma^{W^-} = 52.95$ nb (CMS), 22.88 nb (ALICE).
}
\label{sigs}
\end{table}

\begin{table}[htb]
\begin{tabular}{ccccc}
Detector & $N(S=1)$ &  $N(S=S_1)$ &  $N(S=S_2)$ & $N(S=S_3)$ \\ \hline
\multicolumn{5}{c}{$Z^0$} \\
CMS    & $2.94 \times 10^4$  & $2.07 \times 10^4$ & $2.05 \times 10^4$ 
& $2.72 \times 10^4$  \\
ALICE  & $1.26 \times 10^4$ & $8.88 \times 10^3$ & $8.99 \times 10^3$ 
& $1.20 \times 10^4$ \\
\multicolumn{5}{c}{$W^+$} \\
CMS    & $4.59 \times 10^4$ & $3.17 \times 10^4$ & $3.12 \times 10^4$ 
& $4.24 \times 10^4$  \\
ALICE  & $1.98 \times 10^4$ & $1.35 \times 10^4$ & $1.36 \times 10^4$ 
& $1.86 \times 10^4$ \\
\multicolumn{5}{c}{$W^-$} \\
CMS    & $4.86 \times 10^4$ & $3.35 \times 10^4$ & $3.31 \times 10^4$ 
& $4.48 \times 10^4$  \\
ALICE  & $2.04 \times 10^4$ & $1.39 \times 10^4$ & $1.41 \times 10^4$ 
& $1.92 \times 10^4$ \\
\end{tabular}
\caption{Number of vector bosons produced at $b=0$ in a one month ($10^6$ s)
Pb+Pb LHC run at 5.5 TeV/nucleon.  Note that no decay branching ratios have
been included.}
\label{rates}
\end{table}

\end{document}